# Relaxation dynamics in covalently bonded organic monolayers on silicon

*Nicolas Clément, Stéphane Pleutin, David Guérin and Dominique Vuillaume*

Institute for Electronics Microelectronics and Nanotechnology (IEMN),
CNRS, university of Lille.
BP60069, avenue Poincaré, Villeneuve d'Ascq, F-59652cedex (France).
E-mail: (dominique.vuillaume@iemn.univ-lille1.fr ; nicolas.clement@iemn.univ-lille1.fr)

**Abstract.**

We study the dynamic electrical response of a silicon-molecular monolayer-metal junctions and we observe two contributions in the admittance spectroscopy data. These contributions are related to dipolar relaxation and molecular organization in the monolayer in one hand, and the presence of defects at the silicon/molecule interface in the other hand. We propose a small signal equivalent circuit suitable for the simulations of these molecular devices in commercial device simulators. Our results concern monolayers of alkyl chains considered as a model system but can be extended to other molecular monolayers. These results open door to a better control and optimization of molecular devices.

## I. INTRODUCTION

Molecule-based devices are envisioned to complement silicon devices by providing new functions or already existing functions at a simpler process level and at a lower cost by virtue of their self-organization capabilities. Molecular monolayers represent a widely used systems covering a large range of applications such as molecular memories,[1] switches,[2-5] nanodielectrics for organic and carbon nanotube transistors,[6-8] or bio-sensors.[9] The physics and engineering of the molecule/electrode contact has long been recognized as a key issue, and this has been mainly studied through the impact on the static (DC) characteristics of these molecular devices.[10, 11] Here we use admittance spectroscopy to study quantitatively the dielectric response of highly ordered monolayers grafted on silicon in a parallel-plate-capacitor structure with different interfaces and electrodes. We study both the contribution of dipolar relaxation, molecular reorganization in the monolayer, and the contribution of defects at the silicon/molecule interface. We propose a small signal equivalent circuit suitable for the simulations of these molecular devices in commercial device simulators. Our results concern monolayers of alkyl chains considered as a model system but can be extended to other molecular monolayers. These results open door to a better control and optimization of molecular devices.

Dielectric properties of organic monolayers sandwiched between two contact electrodes in a parallel plate structure are a key issue for a complete understanding of the electronic properties of any nanodevices using functional molecules. Up to now, only DC characteristics (e.g. current-voltage curves) were used to access electron transport through the monolayers, and especially to emphasize the role of the molecule-electrode interface.[12-14] The dynamic electron transport properties in molecular junctions have only been studied by inelastic electron spectroscopy to characterize the interaction between electrons and the molecular vibrations in the infra-red region (>$10^{12}$ Hz).[15-17] However, other relaxation phenomena in the intermediate frequency range (1 - $10^6$ Hz) are also important to assess the molecular organization and the presence of defect in the monolayers. Permanent molecular dipoles tend to relax to be oriented with the electric field direction.[18] Their dynamics is expected to depend on the molecular packing and organization (e.g. presence of domains and clusters) in the monolayers. We note that dynamic relaxations were previously characterized for alkyl monolayers grafted on porous glass[19] and on thick-oxide using a lateral electrode configurations.[20] These authors were able to distinguish the dipole relaxation coming from the bottom and top of the organic monolayers[19] and to follow the dependance of these dipole relaxation as a function of the monolayer density.[20] At lower frequency (< 1 kHz), other phenomena regrouped under the name "interfacial polarization" have to be considered. For instance, we recently showed that electrically-active defects can be observed in monolayer through the measurement of the low-frequency noise (< 100 Hz) of the tunneling current,[21] and we showed how they can modify the tunnel current in the molecular junction.

## II. EXPERIMENTAL SECTION

The electric polarization of a solid has several contributions that may be termed intrinsic e.g. electronic, ionic or dipolar polarization, or extrinsic e.g. polarization caused by defects or traps. Those contributions act in different frequency regions and show particular signatures in the dielectric constant or susceptibility (**Figure 1**) as shown in many textbooks.[22] The real part of the response function describes polarization while their imaginary part takes account for dissipation. Here, we used admittance spectroscopy[23] which is a very sensitive measurement technique for probing molecular relaxation processes on a wide range of time scales. This method characterizes molecular scale dynamics through its relationship to transient changes in the dielectric properties of a macroscopic sample. The method is sensitive enough to measure the AC admittance $Y_m = G_m + jC_m\omega$ of a molecular monolayer junction (containing less than $10^{11}$ molecules) in the range 20-$10^6$ Hz (see section II.B) and as function of a superimposed static potential in the range 0 - 1V. $G_m$ and $C_m$ are the measured parallel conductance and capacitance.

**A. Sample manufacturing.**

We prepared high quality alkyl chain monolayers on both oxide-free silicon wafers and on ultra-thin (0.6 nm thick) thermal oxides. This allowed us to compare two types of silicon/ molecule interfaces. On both cases, we used a 18 carbon atom chain (octadecene, OD, on silicon, and octadecyltrichlorosilane, OTS, on oxidized silicon). The chemical structures of the molecules are shown in **figure** 2**.** Samples A and B: Si-C linked alkyl monolayers were formed by thermally induced hydrosilylation of alkenes (octadecene, see fig. 2) with hydrogen-terminated Si surfaces, adapting a procedure described elsewhere.[24] We used n-type doped (0.02 - 0.5 Ω.cm) Si(111). Two categories of devices, referred to as A and B, with different DC tunnel conductance $G_T$ (low and high respectively) are selected with the aim to evaluate its impact on the dielectric response. These two types of samples are selected (after DC current-voltage measurements) from the same batch as a result of the inherent dispersion already observed with such a chemical approach. Sample C: a 2 nm thick oxide was thermally grown on n-type doped (1-3x$10^{-3}$ Ω.cm) Si(100) in dry $0_2$ at 720°C and chemically thinned to 0.6 nm at 0.64 Å/min by a HF(0.1%)/HCl(2.5%) solution. A monolayer of octadecyltrichlorosilane (OTS) -see Fig. 2- was formed by reacting in solution the freshly prepared oxide surface with $10^{-3}$ M of OTS in a mixture of hexadecane/$CCl_4$ as reported elsewhere.[25, 26] In that case the alkyl chain is bonded to the substrate through siloxane (Si-O-Si) bonds. The quality of the monolayers was assessed by spectroscopic ellipsometry and water contact angle measurements (table 1). We selected only monolayers with a water contact angle larger than 110° and a thickness close to $d_0 cos\theta$ (within ± 3 Å) where $d_0$ is the expected length of the molecule in its all-trans conformation and $\theta$ is the tilt angle between the long axe of the molecule and the normal to the surface (~ 15° and ~ 30° for OTS on oxidized Si and OD on Si:H, respectively).[27, 28] We also prepared a control experiment sample (no organic monolayer) with a 3 nm thick thermal oxide using the same oxidation process as for sample C (dry $O_2$ at 720°C). A 50 nm thick aluminium contact pads with different surface areas between $9x10^{-4}$ cm$^2$ and $4x10^{-2}$ cm$^2$ were vacuum ($10^{-7}$ Torr) evaporated at 3 Å/s on top of the alkyl chains, or they were contacted with a hanging Hg drop. In case of evaporated metal electrodes, the devices were contacted with a micro-manipulator probe station (Süss Microtec PM-5) placed inside a glove box (MBRAUN) with a strictly controlled nitrogen ambient (less than 1 ppm of water vapor and oxygen). The hanging naked Hg drop was formed using a controlled growth mercury electrode system (BAS) in a glove box, the sample was placed on an elevator stage and put into contact by moving it upward. The junction size was estimated by measuring the Hg drop profile by a calibrated USB webcam. The Hg drop contact method has been demonstrated to be shortage free and reproducible.[11, 29, 30] All electrical measurements were conducted in dark. Some OTS samples were submitted to a thermal annealing at 200°C in $N_2$ for 1 hour. We measured about 20 samples of each type to assess the dispersion of the junctions parameters, the curves shown in the figures are representative of the average behavior.

**B. Admittance spectroscopy measurements.**

The DC voltage bias (0 to 1V) superimposed with a small AC signal (10 $mV_{eff}$) was applied on the metallic electrode, the Si substrate was grounded, and the complex admittance was measured using an impedance-meter Agilent 4284A in the range 20-$10^6$ Hz. Measured conductances and capacitances are corrected from a small series resistance $R_S$ (typically in the range 5 - 70 Ω for our experimental set-up) according to standard procedures.

[31] The Agilent 4284A provides directly the parallel conductance and capacitance values, $G_m$ and $C_m$, which are normalized by the contact surface area. To remove the series resistance, $R_s$, contribution and extract intrinsic dielectric properties, $G_m$ and $C_m$ are corrected using eqs. (1) and (2) to give the corrected conductance and capacitance at each measurement pulsation $\omega$:

$$G_c(\omega) = G_m - R_S C_m^2 \left(1 - R_S G_m\right)\omega^2 \tag{1}$$

$$C_c(\omega) = \frac{1 - \sqrt{1 - \left(2C_m R_S \omega\right)^2}}{2C_m \left(R_S \omega\right)^2} \tag{2}$$

We get a corrected admittance per surface unit $Y_c = G_c + jC_c\omega$, from which we deduced two quantities $C'$ and $C''$ related to the real and imaginary part of the dielectric susceptibility of the monolayer $\chi = \chi'(\omega) + j\chi''(\omega)$ , by

$$C' = C_C - C_\infty = \frac{\varepsilon_0 \chi'}{d} \quad and \quad C'' = \frac{\varepsilon_0 \chi''}{d} = \frac{G_C - G_T}{\omega} \tag{3}$$

where $C_\infty$ is the high frequency part of capacitance due to ionic and electronic contribution (Fig.1), and $G_T$ is the tunneling conductance per surface area (measured from DC current-voltage curves or from $\lim_{\omega \to 0} \mathrm{G_C}(\omega)$ ), d is the monolayer thickness, $\varepsilon_0$ the vacuum permittivity and $\omega$ the measurement pulsation. There is still a small parasitic effect (increase of $C'$ and $C''$ at high frequencies) which is due to the sample environment (cables). It is a known effect,[32] but since it only affects the highest frequency part of the curves, we let it as it is.

## III. THEORY

In the limit of a response that is linear in the applied electric field (weak AC-field), the complex dielectric susceptibility is related to the relaxation function, $\phi$, which is the response function of the system after the abrupt removal of a constant electric field

$$\chi(\omega)=\chi'(\omega)+j\chi''(\omega)=\varepsilon_r(\omega)-\varepsilon_r(\infty)=\left(\chi_0-\chi_\infty\right)\left(1+\int_0^{+\infty}\phi(t)e^{j\omega t}\,dt\right) \tag{4}$$

We are interested in the low frequencies part of the dielectric susceptibility attributed to defect or dipolar relaxation (see Fig. 1). $\chi_0$ is the value at zero frequency and $\chi_\infty = 0$, since at high enough frequencies the mobile dipoles cannot follow the field anymore and do not contribute to the polarization. $\varepsilon_r(\infty)$ is the part of the relative dielectric constant due to ionic and electronic contribution; more precisely, it is defined for intermediate frequencies large enough for the dipoles to be "frozen" and not contributing to the polarization but small enough to not consider dissipation from the ionic and electronic degrees of freedom.

In the simplest case, considered long ago by Debye,[33] the relaxation function is given by a single exponential

$$\phi_{Debye}(t)=e^{-t/\tau} \tag{5}$$

where $\tau$ is the characteristic time of relaxation. This is for instance the relaxation function of a rigid polar molecule in a viscous isotropic fluid. Such simple response yields

$$\chi_{Debye}(\omega)=\chi_0\frac{1+j\omega\tau}{1-\omega^2\tau^2} \tag{6}$$

The real part of Debye's susceptibility shows a plateau at low frequencies below $1/\tau$ and decreases as $\omega^{-2}$ above $1/\tau$. Its imaginary part shows a peak at $1/\tau$: it increases as $\omega$ in the pre-peak region and decreases as $\omega^{-1}$ in the post-peak region. However, this type of response is not usual in condensed matter. Instead, fractional power laws are most often observed,

$$\begin{aligned}\chi(\omega)&\propto(j\omega)^{n-1} \quad for \quad \omega\gg\omega_p\\ \chi(0)-\chi(\omega)&\propto(j\omega)^{m} \quad for \quad \omega\ll\omega_p\end{aligned} \tag{7}$$

$\omega_p$ being the peak of maximum loss. This is the well-known 'universal' dielectric response pointed out by Jonscher.[18] Our data clearly show two contributions (Figs. 3-5). We write

$$\chi(\omega) = \chi_1(\omega) + \chi_2(\omega) \tag{8}$$

where 1 and 2 refer to the first and second (at higher frequencies) peaks as shown in Figs. 3-5 and are discussed below.

**A. Peak 1, interfacial contribution**

The first peak in the dielectric response (Fig. 1) is usually attributed to interface defects.[31] The internal dynamics of such defects, modeled by two level systems, results in a Debye contribution to the dielectric susceptibility[31]

$$\chi_1(\omega) = \frac{q^2 N_T}{k_B T} \frac{1}{1 - j\omega\tau_1} \tag{9}$$

where $N_T$ is the defect density, $1/2\pi f_1 = \tau_1$ with $f_1$ the frequency of the peak 1 maximum, $k_B$ the Boltzmann constant, $T$ the temperature and $q$ the electron charge. The characteristic time $\tau_1$ of relaxation may be due to thermal excitation or interaction with the phonons of the bath, for instance.

**B. Peak 2, molecular contribution**

This contribution usually follows the 'universal' behavior of Jonscher.[34] The response is then characterized by three parameters: the pre- and post-peak exponents, $m$ and $n$, respectively, and the peak position, $f_2$ (as shown in Fig. 1). Several phenomenological expressions have been proposed to mimic the universal dielectric response.[22] A suitable expression that models remarkably well our data is the Dissado-Hill dielectric susceptibility.[35] It reads,

$$\chi_2(\omega) = \chi_0 \frac{\Gamma(1+m-n)}{\Gamma(m)\Gamma(2-n)} \left(1 - j\omega\tau_2\right)^{n-1} {}_2F_1\left(1-n, 1-m, 2-n, \left(1 - j\omega\tau_2\right)^{-1}\right) \tag{10}$$

with $1/2\pi f_2 = \tau_2$ , $f_2$ being the frequency of the peak 2 maximum, and $\Gamma$ the gamma function (*n* and *m* are defined above). Eq. (10) follows the asymptotic limits given by eq. (7). ${}_2F_1$ is the Gauss hypergeometric function with $0 \leq m \leq 1$ and $0 \leq n \leq 1$. Dissado and Hill have proposed this function based on some qualitative arguments. Their function has proven to model successfully numerous data of very different systems. Moreover they have tentatively related the *n* and *m* exponents to structural characteristics. Derived from the generic arguments they have developed we propose below a qualitative model to understand dissipation in our device.

## IV. RESULTS

The **figures 3 and 4** show *C'* and *C"* for two representative samples hereafter denoted as samples A and B. They are both made of OD on Si but differ by their level of measured DC tunnel current (see above, section II.A). We distinguish 3 contributions corresponding to peaks in the dielectric loss *C"* and steps and plateaus in the capacitance *C'*. The first peak in *C"* (filled in yellow) corresponds to interfacial relaxations and the second peak (in blue) to molecular relaxations. The contribution of the DC tunnel conductance is filled in green (it follows a *1/f* behavior, $G_T/\omega$ contribution in eq. (3)). In all curves, we observe in the high frequency part an increase of both *C'* and *C"*, which is due to a parasitic inductance (see section II.B). The first peak is well fitted with the Debye model[33] describing a set of identical non-interacting defects whose dynamics is reduced to tunneling between two minimal configurations (see section III on theory). It can be linked to the bias dependent *1/f* noise increase due to a peaked density of defect previously reported for this device.[21] The amplitude of this peak varies from device to device: the interfacial relaxation contribution is relatively weak for sample A and strong for sample B. We note an increase of the *C"* peak and the *C'* plateau by about 30 times. This increase is strongly related with an increase in the DC tunnel conductance $G_T$ by 3 orders of magnitude. Device B was selected (see section II.A) to illustrate this point. The second peak was fitted with a Dissado-Hill expression[35] (see section III). The imaginary part of this phenomenological expression shows fractional power laws, $\omega^m$ and $\omega^{n-1}$, in the pre and post peak region, respectively. Such behaviors are the signature of strongly constrained dynamic of relaxation due to interaction between the relaxing quantities and their environment. The *n* and *m* exponents have been related by Dissado and Hill to structural characteristics. The dipolar relaxation contributions to *C'* and *C"* are similar for samples A and B, which means that they have a similar molecular organization and packing. Since the nature and technology used to form the top contact on the monolayer can strongly influence and even degrade the electronic properties of the molecular junction,[11, 36] we also tested samples A and B using a hanging Hg drop. The results (Fig 4) are almost similar as for the evaporated Al electrode. The dipolar relaxation contributions to *C′* and *C″* have almost the same amplitude, however, the interfacial peak position is shifted by the difference in work function between Al and Hg (see later, discussion section).

The interface with the Si is also playing an important role. The presence of an ultra-thin oxide layer (0.6 nm) between the molecule and the Si electrode electronically decouples the molecule from the Si, while a stronger electronic coupling is obtained without the oxide. These features induce changes in the DC current-voltage characteristics.[37] **Figure 5** shows the measured *C′* and *C″* for sample C (OTS on a slightly oxidized Si, see section II.A). The interfacial relaxation is still observed with almost the same amplitude as for sample A. This means that this contribution comes from the molecule and is not strongly related to the nature of the substrate (no peak from oxide as checked on a reference Si/$SiO_2$ sample). Finally, the sample C was annealed at 200°C under $N_2$ atmosphere. We observed a strong decrease of the interfacial relaxation peak (Fig. 5).

# V. DISCUSSIONS

## A. Peak 1, interfacial contribution

According to the above considerations and previous literature on other semiconductor-insulator-metal results,[31] we assign the interfacial relaxation contribution (peak 1) to defects, traps, localized at, or near, the interface between the monolayer and the substrate. This hypothesis can be considered for several reasons: (i) we have already pointed out the link between interface-trap assisted tunneling current and low frequency noise in these molecular junctions,[21] considering traps energetically localized in the HOMO-LUMO gap of the alkyl monolayers; (ii) interface traps in other inorganic tunnel barrier usually respond in this frequency range.[31] We note, however, that we did not observe (Fig. 6) a significant interfacial relaxation peak in this frequency range for a reference sample (i.e. a 3 nm thick thermal oxide without any molecular monolayer, see section II.A), which means that the observed peaks are due to some defects related to the organic monolayer and not at the Si/$SiO_2$ interface, nor buried in the $SiO_2$ layer. The nature, origin, of which remains to be discovered. One possible origin should be interface defects at the molecule-substrate interface (e.g. dangling or distorted bonds).[38, 39] Because covalently bonded organic monolayers are organized in domains with grain boundaries and they present also pinholes, another possible origin could be surface states coming from substrate surface atoms belonging to these regions where molecules are absent. It has been previously observed and calculated that these pinholes can play an important role to determine the overall electrical behavior of the monolayer devices.[40, 41] We can note that the fact that the first contribution is very close to a Debye response is attributed to a distribution of defects with a time constant response strongly peaked around the frequency $f_1$: one type of defects seems to dominate the dielectric response in this frequency range. We can deduce the trap density, $N_T$, from the following equation:[23]

$$C' = C_\infty + \frac{C_D}{(1 - j\omega\tau_1)} = C_\infty + \frac{q^2 N_T}{k_B T} \frac{1}{(1 - j\omega/2\pi f_1)} \tag{11}$$

where $C_\infty$ is high-frequency limit of $C'$, $C_D$ is the capacitance associated to charges trapped on defects, $\tau_1$ is a characteristic time constant of these traps, $T$ is the temperature, $k_B$ the Boltzmann constant, $q$ the electron charge and $f_1$ the frequency at the maximum of the peak 1. By doing the above measurements at various DC bias, the Fermi energy of the Si is swept along the energy distribution of the traps. **Figure 7** plots the corresponding trap density distribution for the samples A and B (with Al and Hg drop contacts). Sample A (with both Al and Hg electrodes) displays a remarkably low density (in the range $1.2\text{x}10^9$ - $10^{10}$ cm$^{-2}$) for a room-temperature process, almost on a par with the state-of-the art Si/$SiO_2$ interface, but, in this latter case, obtained after a high-temperature post-oxidation annealing. Samples B (both Al and Hg electrodes) have a higher density of traps ($2\text{x}10^{10}$ - $10^{12}$ cm$^{-2}$), and we clearly observed a shift in the $N_T$-$V_{DC}$ characteristics due to the difference in work functions between the Al and Hg electrodes. This means that the defects are located near the alkly/Si interface since they are sensitive to the Si surface potential.[31] For $V_{DC} > 0$, the probed traps are located at energy close or slightly higher than the Si conduction band since the Si is in accumulation regime.

**B. Peak 2, molecular contribution**

The second peak can be linked to molecular relaxation. We found that the Dissado-Hill function (see above, eq. 10) is appropriate to fit both $C'$ and $C''$. It is characterized by three parameters: the pre and post peak exponents, $m$ and $n$, respectively, and the peak position (see Fig. 1). Table 2 summarizes these parameters (at $V_{DC}$=0.2 V), and the maximum peak 2 frequency, $f_2$, deduced from the fits for the three devices shown in Figs. 3-5. According to ref. [35], $n$ is ascribed to reflect the order within clusters ($n$ = 1 representing a perfectly ordered cluster) and $m$ is ascribed to characterize cluster size fluctuations ($m$ = 0 representing a completely frozen – no fluctuation – structure). The alkyl chains have two small permanent dipoles localized at both ends but most of the molecules are strongly constraint since they are sandwiched between the substrate and the top electrode. The amplitudes of the two local permanent dipoles may be roughly estimated by semi-empirical calculations between 0.5 and 1 D (MOPAC software).[42] We did the calculation by considering a single molecule attached to a silicon cluster (88 Si atoms, see Fig. 2-c). These values are clearly an overestimation since screening effects from the metallic electrodes and interactions between molecules have not been considered. In the case of OTS on oxidized silicon, the dipole is a little bit larger (1.5-2 D) since the Si-O bond is more polar than the Si-C bond. The presence of these permanent dipoles, even small dipoles, should be sufficient to give sizable dielectric polarization but, in our case, a thin electrode is deposited on top of the monolayer. As a result most of the molecules are strongly constraint and can hardly react to an electric field: they can be considered as inactive during the polarization process. However, we assume that some molecules - or at least segments of some molecules - are more free. They may be preferentially located at (or near) some structural defects of the monolayer. The local permanent dipoles of those molecules are then mobile, or partially mobile, and are consequently able to follow the applied electric field. They give the most important contribution to the dipolar polarization. Moreover, when they are reoriented by the applied field they affect also the rest of the monolayer due to the van der Waals type of interaction that exist between molecules. The reorientation of the mobile dipoles and the disturbance that these motions induce in the surrounding of the mobile dipoles are both responsible for the observed dielectric susceptibility. In this picture, the details of the dielectric relaxation are due to the interaction of the active dipole with the harmonic modes of the monolayer. They should be sensitive to the way the monolayer mechanically reacts to the local constraints caused by the reorientation of the active dipoles. The measured response is a system average which contains elements both from the active dipoles and from the mechanical interaction of these dipoles with the rest of the structure. The basic case is OTS SAM (sample C). $m$ = 0.1 - 0.2 and $n$ = 0.8 - 0.9 exponents are close to the flat loss limit ($m$ = 0 and $n$ = 1). These are typical values for imperfectly crystallized materials and it is well-known that OTS SAM are polycrystalline with a short-range order (coherence length of 70 Å from X-ray grazing angle diffraction). [43] We estimate (MOPAC software)[42] the small dipole at the foot of the molecule in the range 0.5-2 D (depending on the OD or OTS samples, see above) when the alkyl chain is in all-trans conformation and a little higher, if we include a gauche-defect somewhere in the chain conformation. Such a local dipole has been also suggested to explain the dielectric loss measurements done on alkyl chains grafted on porous glass[19] and thick-oxide.[20] A more quantitative comparison of the peak amplitude and frequency remains difficult given the wide differences in the experimental conditions (low temperatures and disordered, low packing, monolayers in their cases, room temperature and highly-packed, better-ordered, monolayer in our case) and differences in the sample geometries (porous glass and planar electrodes for the quoted works, vertical structure with a sandwiched monolayer here).

For OD SAM (samples A, B), the $m$ exponent is found to be more than two times larger ($m$ = 0.2 - 0.45) at low electric field. The molecular packing (the density of molecules is two times higher for OTS than OD) and the molecular organization (e.g. the tilt angles of the long hydrocarbon chain with respect to the surface normal, $\theta_{OTS}$ = 15°, $\theta_{OD}$ = 34°) differ for the two types of monolayers.[27, 28] Therefore, we expect changes in the interaction between the active dipoles and the inactive part of the SAM that should result in different exponents. These variations of $m$ seem in agreement with the statement of Dissado and Hill mentioning that the more the structure of the system is mechanically rigid the more $m$ approaches 0. This is corroborated by the behavior of $m$ versus the application of a constant electric field (Fig. 7 for sample A with Al electrodes). The dipole fluctuations are reduced ("frozen") by increasing a DC field and therefore *m* should be reduced as observed in Fig. 7. The peak frequency is also shown to be linearly shifted to high frequencies. This can be explained in simple terms assuming the dynamics of the mobile dipoles described by a damped harmonic oscillator (see appendix A).

To summarize:

i) *m* depends on the nature of the monolayer. Since the density of molecules is larger for OTS than OD monolayer, we expect the OTS monolayer to be more rigid and therefore $m_{OTS} < m_{OD}$. This is indeed what we have observed (table 2 and figures 3-5).

ii) *m* is changed when a DC electric field is applied perpendicularly to the electrode surface. We expect the continuous field to rigidify the structure and therefore $m_{E\neq 0} < m_{E=0}$. This is the behaviour that we have observed (Fig. 7-b).

## VI. ELECTRICAL EQUIVALENT CIRCUIT.

We deduced a small signal equivalent circuit for these Si-monolayer-metal junctions (**Fig. 8**). A perfect molecular tunnel junction would be modeled with a tunnel conductance $G_T$ in parallel with the capacitance $C_\infty$. The admittance for the Debye peak (interfacial relaxation, peak 1) is $\mathrm{Y_D} = \mathrm{jC_D}\omega/(1+\mathrm{j}\omega\tau_1)$, which corresponds to a capacitance $C_D$ in series with a conductance $\tau_1/C_D$, where $\tau_1$ and $C_D$ are defined and given by eq. (11). The specific admittance for the molecular relaxation (peak 2) is deduced from the Dissado-Hill function as:

$$Y_{mr} = \frac{jC_{mr}(V)\omega}{(1-j\omega\tau_2)^{1-n}}\,{}_2F_1\left(1-n, 1-m, 2-n, \frac{1}{1-j\omega\tau_2}\right) \qquad (12)$$

where ${}_2F_1$ is the Gauss hypergeometric function (see also eq. 10), and the parameters *n*, *m*, $C_{mr}$ and $\tau_2 = (2\pi f_2)^{-1}$ are given from the fits of the molecular relaxation peak (table 2). The quantitative contribution of each of these loss mechanisms on the total capacitance $C'$ is obtained by comparison of $C_D$ and $C_{mr}$ with $C_\infty$. It is clear that the contribution of $C_D$ to $C'$ is weak for device A (<10 % of $C_\infty$ ,Fig. 3-a, table 2), but very large for device B (up to 300 % of $C_\infty$ at $f$ < 100 Hz, Fig. 4-a, table 2). Preliminary studies on technological issue (thermal annealing, assembly process) let us think that it can be reduced by thermal annealing (see Fig. 5). The contribution of molecular relaxation $C_{mr}$ to the total capacitance is weak but non negligible (~10 % $C_\infty$ at 10 Hz, low $V_{DC}$). It is the dominant mechanism of loss for high quality junctions (low $C_D$), as observed for device A at low $V_{DC}$. However this contribution would dominate in the case of high-k nanodielectrics[8] using push-pull molecules with a larger dipole.

## VII. CONCLUSIONS

In summary, we have shown how to distinguish between the relaxation dynamics of molecules and the dynamics of molecule/electrode interface by admittance spectroscopy. The interfacial relaxation is related to defects found in Si/SAM or Si/$SiO_2$/SAM molecular junctions (not in Si/$SiO_2$). They are located at the bottom of the SAM and at energy close to the conduction band edge of the silicon. Depending on their density, DC electrical characteristics are more or less affected. Preliminary studies on technological issues, such as thermal annealing, let us think that this interfacial relaxation peak can be strongly reduced. These results also open the way for a better tuning of the dielectric response of molecular junctions by choosing and designing appropriate selection of molecules.

## ACKNOWLEDGEMENTS

DG thanks David Cahen and Oliver Seitz for their welcome during his stay at the Weizmann Institute. We thank David Cahen and Oliver Seitz for sample preparations (alkene on Si), advise on making a high-quality alkene monolayer on Si-H surfaces and discussions. We thank Stéphane Lenfant for a careful reading of the manuscript. This work was partly supported by the ANR contract # 05-NANO-001-01

## APPENDIX A

When a DC electric field is applied, a linear shift of the maximum loss frequency is observed ($f_2$ in Fig. 6-b). This can be explained in simple terms. We consider a mobile dipole, $\mu$, with disordered orientation, $\phi_0$, that can be dynamically reoriented $\phi(t) = \phi_0+\delta\phi(t)$. The dipole is trapped in a potential assumed to be harmonic $H_{trap} = k\delta\phi(t)^2/2$, with $k$ the elastic constant. It interacts with the electric field, $E_{AC}$ and $E_{DC}$, the AC and DC components, $H_{int} = -\mu\cos\phi(t)(E_{AC}(t)+E_{DC})$. The dissipation is introduced in the simplest way via a constant friction coefficient, $\eta$, that models the interaction with the bath, $H_{dis} = \eta\delta\dot{\phi}(t)$ . After an expansion of the interaction with the field in $\delta\phi(t)$, the equation of motion reads

$$J\delta\ddot{\phi}(t) + \eta\delta\dot{\phi}(t) + \left(k + \mu\cos\phi_0 E_{DC}\right)\delta\phi(t) = \xi(t) + \gamma E_{AC}(t) \tag{A1}$$

where $J$ is the inertial moment of the dipole, $\gamma = \mu\sin\phi_0$ and $\xi(t)$ is a random field related to the frictional coefficient by $\langle \xi(t)\xi(t')\rangle = k_B T\eta$ (< > represents the mean over thermal noise), $T$ the temperature and $k_B$ the Boltzmann constant. In this equation we have replaced the initial dynamical variable by a shifted variable

$$\delta\phi \rightarrow \delta\phi + \frac{\gamma}{k + \mu E_{DC}\cos\phi_0} \tag{A2}$$

Eq. (A1) is the usual classical Langevin equation that can be solved by Laplace's transform. Assuming the dipole dynamics to be overdamped ($J\rightarrow 0$), the susceptibility is readily found to be of the Debye type with a characteristic frequency increasing with $E_{DC}$

$$f = \frac{1}{\tau} = \frac{k + \mu \cos\phi_0 E_{DC}}{\eta} \qquad \text{(A3)}$$

To get a more general susceptibility of the Jonscher type it is necessary to consider more complex friction that depends on time instead of the simple memory less constant taken here. [44] However, with more general friction term the characteristic frequency of the mobile dipole remains unchanged (Eq. A3).

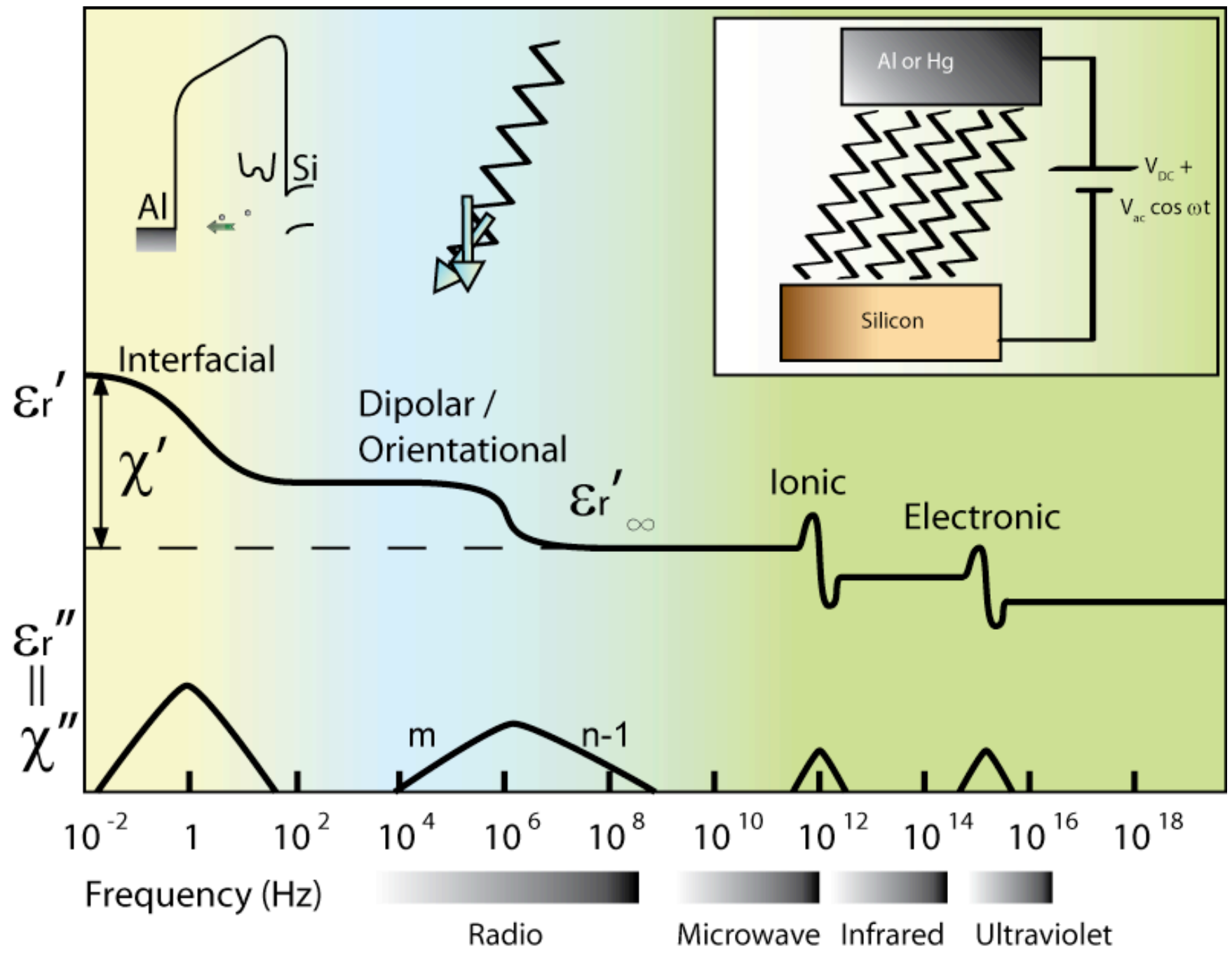

**Figure 1.** Frequency dependence of real, $\varepsilon_r$' (or $\chi$'), and imaginary, $\varepsilon_r$" (or $\chi$"), parts of the permittivity in the presence of interfacial, orientational, ionic and electronic polarization mechanisms. Inset: left: a band diagram with a 2 level tunneling system illustrate the interfacial polarization mechanism; middle: the small dipole linked to each alkyl chain tends to orient with the electric field. It creates a dipolar polarization; right: schematic view of the experimental setup, parallel plate capacitor structure, on Si-n/alkyl chain SAM/Al or Hg molecular tunnel junction.

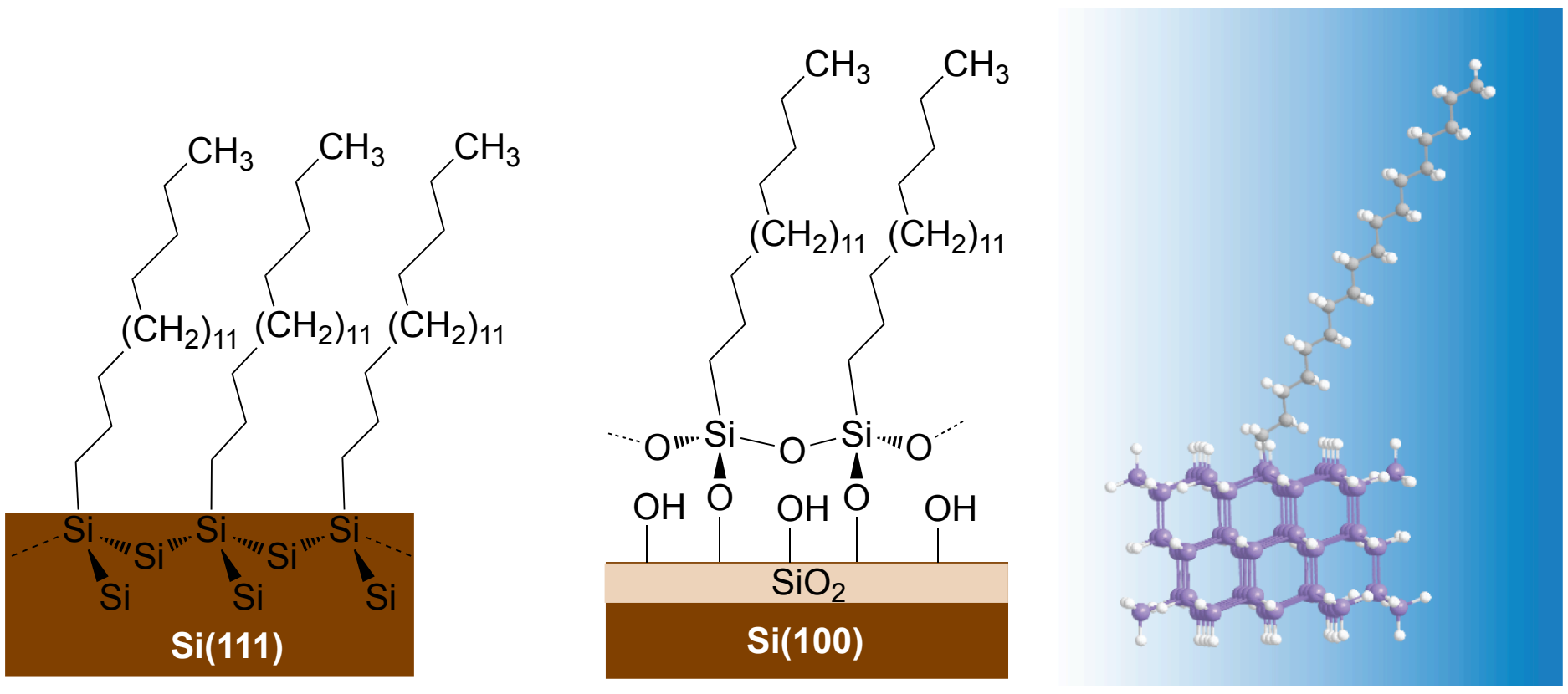

**(a)** **(b)** **(c)**

**Figure 2.** Schematic representation of (a) octadecene (OD) molecules grafted on Si substrate, (b) octadecyltrichlorosilane (OTS) on oxydized silicon substrate, (c) stick and bowl model of the molecule and a Si cluster.

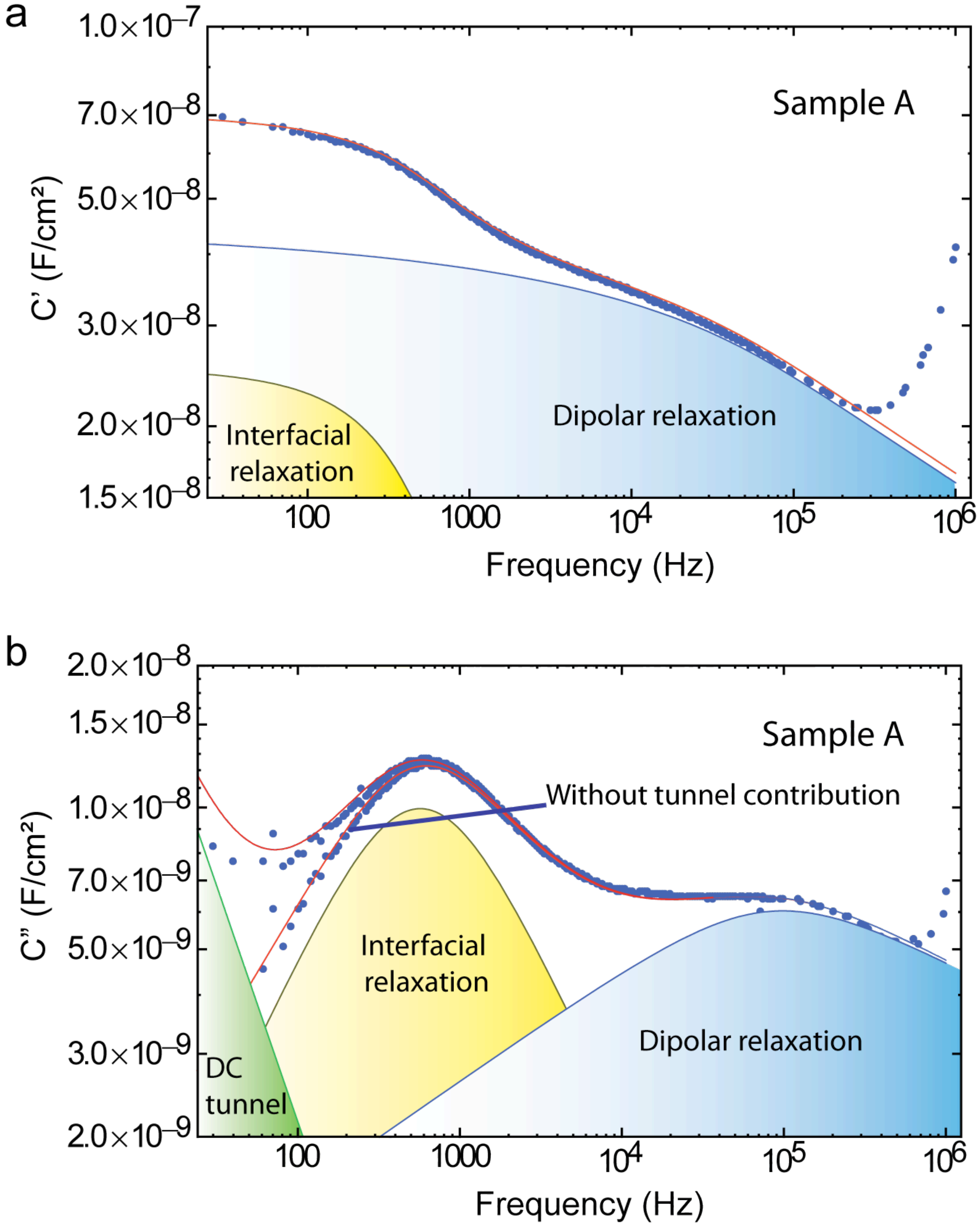

**Figure 3.** $C' \propto \chi'$ and $C'' \propto \chi''$ graphs with frequency dependence for samples A (alkene SAM with low density of interfacial defects **(a-b)**. Experimental data are the dots, with and without the DC tunnel contribution. Fits (lines) include a contribution from interfacial relaxation and dipolar relaxation as indicated. The DC tunnel conductance contribution (plotted as $G_T/\omega$, see eq. (3)) is also represented in $C''$ graphs. The DC bias voltage is $V_{DC} = 0.2$ V.

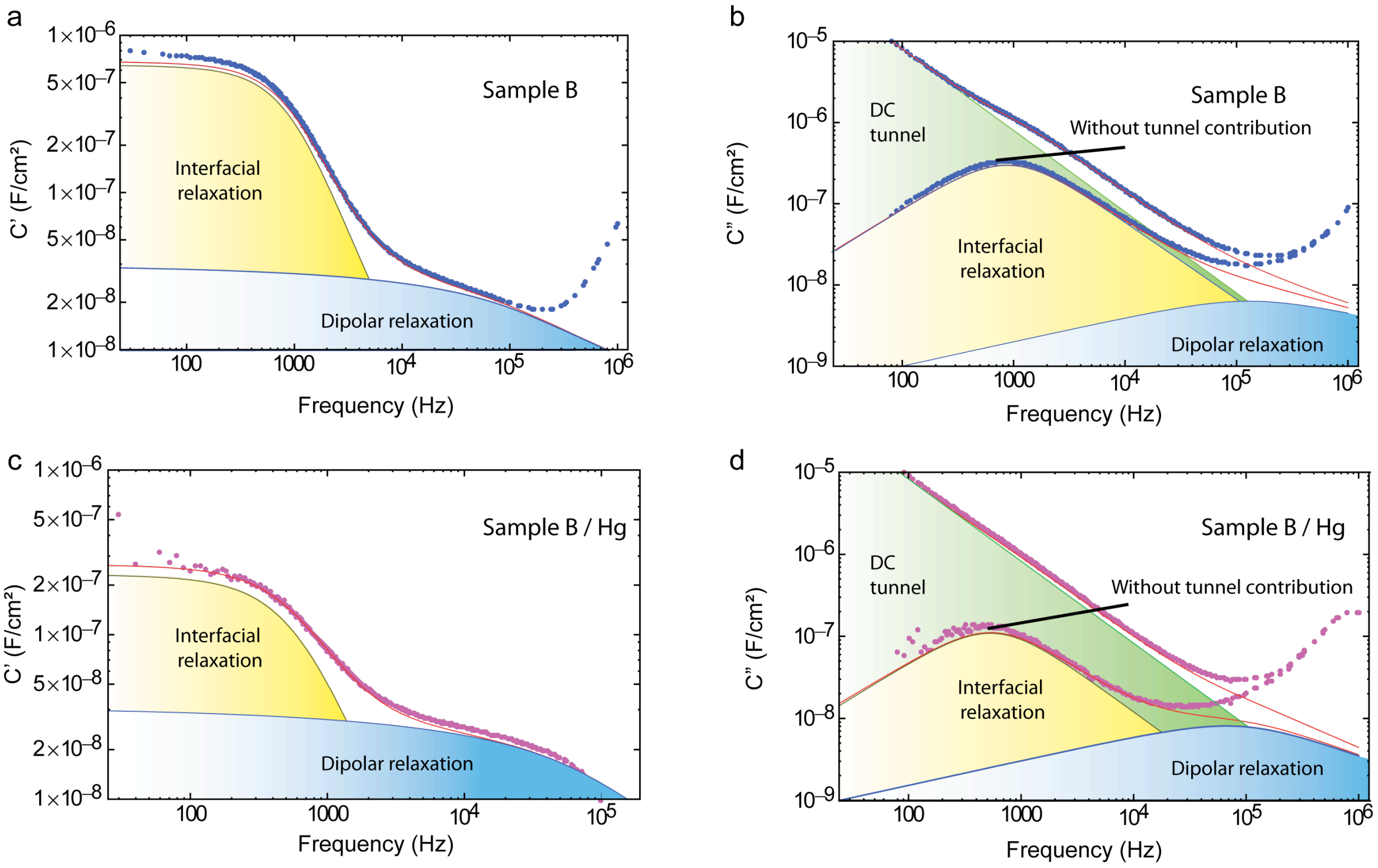

**Figure 4.** Same as figure 3 for sample B (alkene SAM with a large density of interfacial defects contacted by an Al electrode) **(a-b)**, sample B contacted by a Hg drop **(c-d).** The DC bias voltage is $V_{DC} = 0.2$ V

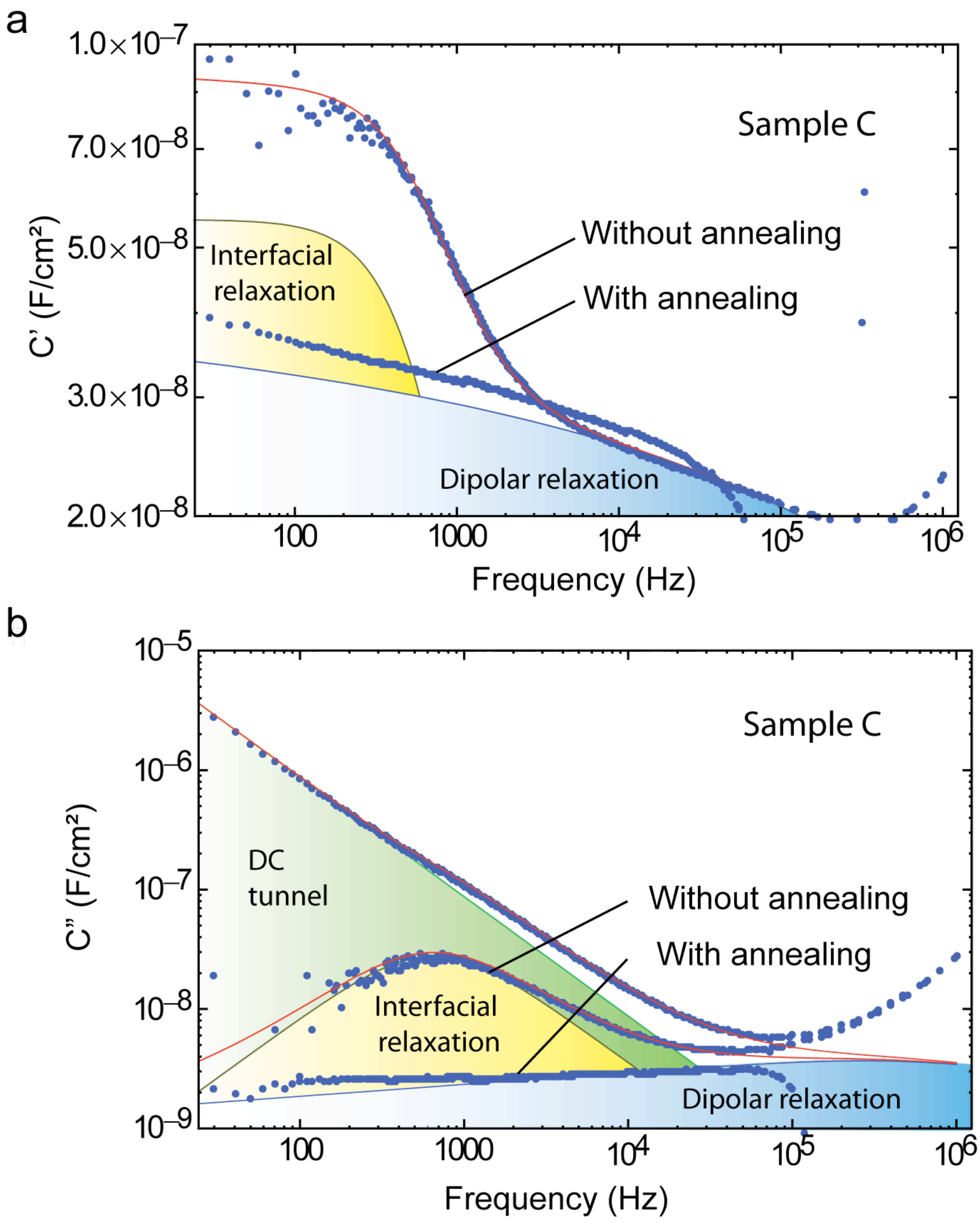

**Figure 5.** Same as figures 3 and 4 for and sample C (OTS SAM with / without annealing).

The DC bias voltage is $V_{DC} = 0.2$ V

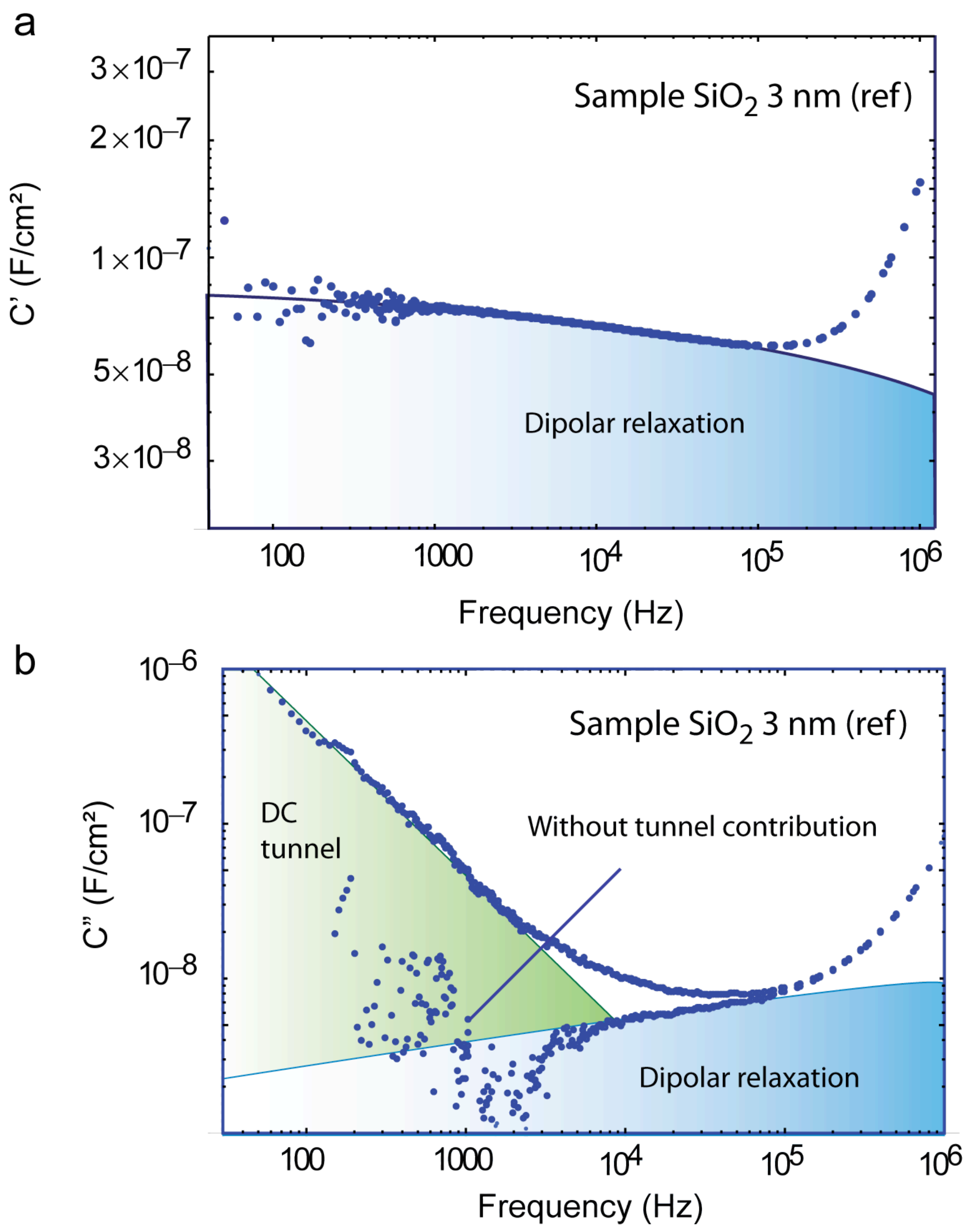

**Figure 6.** Same Figs. 3-5 for the reference samples (3 nm thick $SiO_2$, no SAM). No interfacial relaxation (peak 1) can be detected. The DC bias voltage is $V_{DC} = 0.2$ V

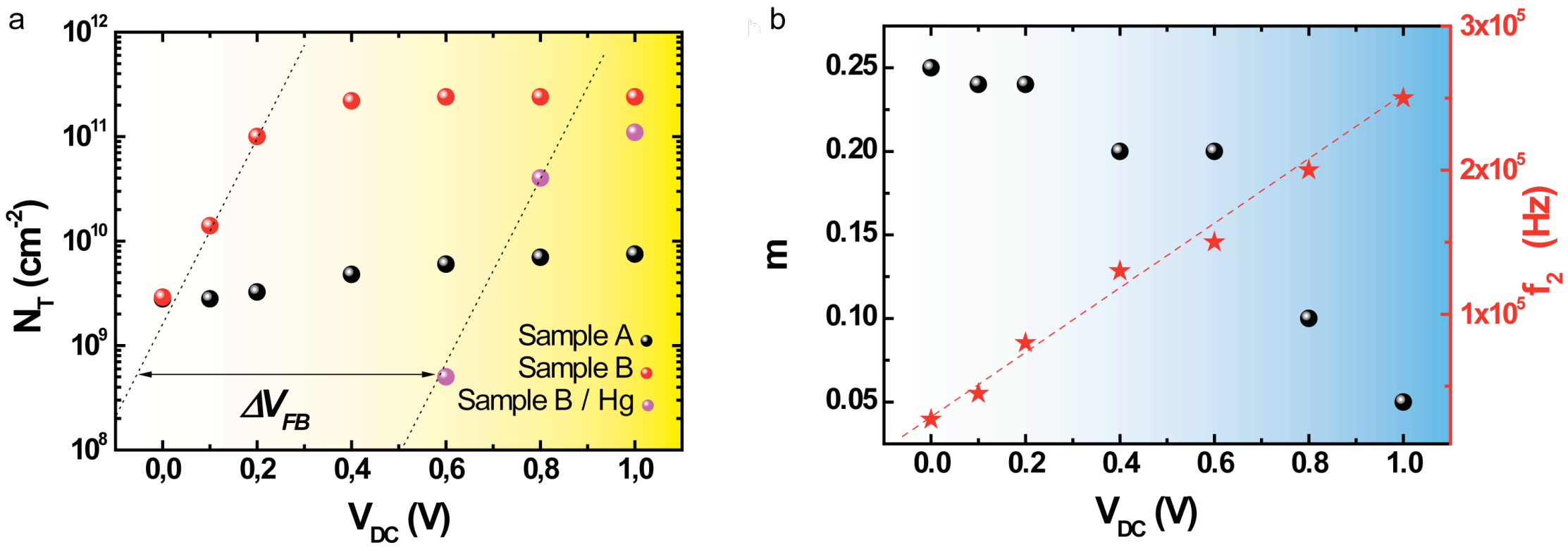

**Figure 7. (a)** The density of defects $N_T$, (eq. 11) as a function of $V_{DC}$ for alkene SAM (Sample A with Al and Hg electrodes, B with Al electrode and Hg electrodes). Curve A(Hg) and B (Hg) are shifted by $V_{FB}$ corresponding to the difference of metal work function between Al and Hg. **(b)** parameters (sample A/Al) for molecular relaxation m (slope before peak) and $f_2$ (frequency of molecular relaxation peak) are plotted as a function of $V_{DC}$.

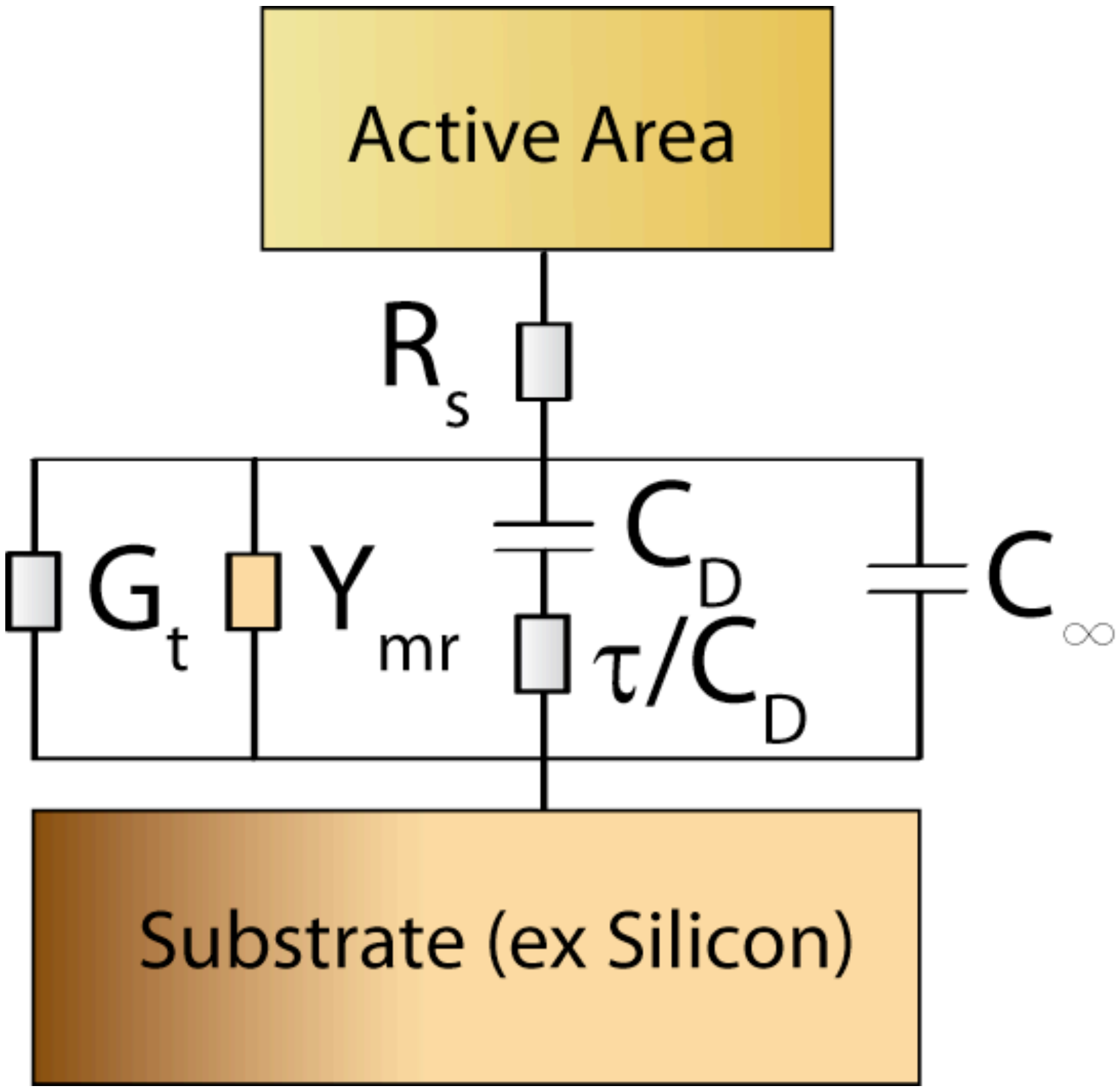

**Figure 8 .** Small signal equivalent circuit for Si / alkyl chain self-assembled monolayers / metal junctions.

**Table 1.** Water contact angle ($\theta_{H20}$), measured thickness ($d$) and theoretical, expected thickness for a monolayer with molecules in the all-trans conformation tilted by an angle $\theta$ with respect to surface normal ($\theta$ ~ 15° and 30° for OTS on slightly oxidized Si (sample C) and OD on Si:H (samples A and B), respectively)

| samples | $\theta_{H20}$ (°) | d (nm) | $d_0\cos(\theta)$ (nm) |
|---|---|---|---|
| A & B | 111 (±1) | 2.18 (±0.15) | 2.2 |
| C | 109 (±1) | 2.50 (±0.15) | 2.6 |

**Table 2.** Fitted parameters (eq. 12) $m$, $n$ and $C_{mr}$ (data measured at low bias, $V_{DC}$ = 0.2 V) and frequency $f_2$ of the dipolar relaxation peak. Parameters $C_D$ for peak 1 (eq. 11) and $C_\infty$ (eqs. 3 and 11). Sample A: alkene on Si (low density of interfacial defects), B: alkene on Si (high density of interfacial defects), C: OTS on Si/$SiO_2$ (6 Å of $SiO_2$). We did not see any significant variation whether the top electrode is evaporated Al or hanging Hg drop, so parameter range given in the table take both cases into account. n.a. stands for non applicable. (*) This value may be underestimated due to the upper limit (1 MHz) of our apparatus.

| Device | A | B | C | ref $SiO_2$ |
|---|---|---|---|---|
| m | 0.2-0.3 | 0.2-0.45 | 0.1-0.2 | 0.15 |
| n | 0.7-0.8 | 0.7-0.8 | 0.8-0.9 | n.a. |
| $f_2$ (kHz) | 80-200 | 60-200 | 370 | $10^3$ (*) |
| $C_{mr}$ (F/cm$^2$) | 5-6 x$10^{-9}$ | 5-6 x$10^{-9}$ | 3-4 x$10^{-9}$ | 9.4x$10^{-9}$ |
| $C_D$ (F/cm$^2$) | 0.6-4.5 x $10^{-8}$ | 0.2-1.5 x $10^{-6}$ | 0.3-2.5 x $10^{-7}$ | n.a. |
| $C_\infty$ (F/cm$^2$) | 4-6 x $10^{-7}$ | 3-4 x $10^{-7}$ | 3-4 x $10^{-7}$ | 5 x $10^{-7}$ |